\documentclass{aa}

\usepackage{graphicx}
\usepackage{txfonts}
\usepackage{booktabs}
\usepackage{hyperref}
\usepackage{amsmath}
\usepackage{mathrsfs}
\usepackage{amssymb}
\usepackage{mathtools}
\usepackage{enumitem}
\usepackage{verbatim}
\usepackage{tikz}
\usetikzlibrary{shapes.geometric, arrows.meta, positioning}
\usepackage{subcaption}
\usepackage{float}
\usepackage{multirow}

\usepackage{natbib}
\bibpunct{(}{)}{;}{a}{}{,}

\begin{document} 

\title{Predicting coronal mass ejection travel times \\ using enhanced model-guided machine learning}

\author{M. Lampani\inst{1}
          \and
          M. Rossi\inst{2}
          \and
          S. Guastavino\inst{1,3}
          \and
          M. Piana\inst{1,3}
          \and
          A.M. Massone\inst{1,3}
          }

\institute{MIDA, Dipartimento di Matematica, Università di Genova, via Dodecaneso 35 I--16146 Genova, Italy\\
\email{sabrina.guastavino@unige.it}\and
Department of Mathematics, University of Pisa, Largo Bruno Pontecorvo 5, 56127 Pisa, Italy\and
Osservatorio Astrofisico di Torino, Istituto Nazionale di Astrofisica, Strada Osservatorio 20, 10025 Pino Torinese, Italy}

\date{Received --; accepted --}

\abstract 
{Coronal mass ejections (CMEs) are key drivers of space weather events, posing risks to both space-borne and ground-based systems. Accurate prediction of their arrival time at Earth is critical for impact mitigation. To this end, physics-informed artificial intelligence (AI) approaches have proven more effective than purely data-driven or physics-based methods, generally offering higher accuracy and better explainability than the former and lower computational cost than the latter. In this work, we propose a generalization of the physics-driven AI framework based on the classical drag-based model (DBM) by integrating the extended version of the drag-based model (EDBM). This enhancement allows us to include in the training process CME events whose interplanetary dynamics are incompatible with those assumed by the DBM. We achieve travel-time prediction accuracy comparable to state-of-the-art methods. We also perform a parametric robustness analysis, highlighting the stability of our approach under small variations in the drag coefficient. Furthermore, we propose a categorization of CMEs into speed regimes defined by the EDBM using a multiclass classification model based on logistic regression, which could be implemented in near-real-time operational space weather forecasting systems. The results show that the EDBM framework broadens the applicability of forecasting models while preserving good predictive accuracy.}


\maketitle

\section{Introduction}
\label{sec:intro}
Coronal mass ejections (CMEs; \citealp{chen2011coronal,howard2014space}; see also \citealp{webb2012coronal} for a review) are large-scale eruptions of magnetized plasma from the Sun’s outer atmosphere into space. These events can release vast amounts of solar material that propagate through the heliosphere. They are closely related to the Sun’s activity cycle \citep{webb1994solar}, which governs the frequency and intensity of solar phenomena. During the solar maximum (the peak of the cycle), CMEs occur more frequently and tend to be more powerful due to increased magnetic activity on the Sun \citep{st2000properties,gopalswamy2005coronal,gopalswamy2006pre}. 

CMEs are primarily detected using a combination of in situ and remote-sensing instruments, which together provide measurements of their kinematic and physical parameters. Coronagraphic observations yield estimates of the CME geometry, initial front speed at eruption, propagation direction, and mass. In situ spacecraft record temporal and compositional variations in the surrounding interplanetary medium, particularly in the solar wind within which the CME is embedded as it travels through space. Remote-sensing detections are provided by instruments located either on orbits around the Sun-Earth Lagrangian point $L_1$, e.g., the Large Angle and Spectrometric Coronagraph (LASCO; \citealp{brueckner1995large}) aboard the Solar and Heliospheric Observatory (SOHO; \citealp{domingo1995soho}), or positioned ahead of and behind the Sun-Earth line, as in the case of the Sun Earth Connection Coronal and Heliospheric Investigation (SECCHI) suite aboard the twin STEREO spacecraft (STEREO‑A and STEREO‑B). In situ measurements are typically carried out near $L_1$ by scientific satellites such as Wind \citep{wilson2021quarter}, the Advanced Composition Explorer (ACE; \citealp{stone1998advanced}), and SOHO's Charge, Element, and Isotope Analysis System (CELIAS; \citealp{hovestadt1995celias}).

The interaction of CMEs with Earth’s magnetic field is a major driver of space weather events (e.g., \citealp{gonzalez1994geomagnetic,pulkkinen2007space,gopalswamy2016history}; see, e.g., \citealp{chernogor2025global} for the recent emblematic case of the geomagnetic storm of 10-11 May 2024). When CMEs are directed toward Earth, they can significantly impact satellites, communication systems, and ground-based infrastructure. Specifically, they can trigger geomagnetically induced currents in power grids, cause signal scintillation affecting GPS and radio communications, disrupt satellite reception, generate geoelectric fields, and even pose radiation risks to astronauts in space. So far, prevention remains the only effective strategy to protect infrastructure from the impacts of CMEs. This underscores the need for accurate and timely predictions of CME arrival times at Earth. Such predictions are essential for implementing precautionary measures, such as powering down vulnerable systems or reorienting satellites, to mitigate potential damage.

However, this task has always proven challenging, mainly due to the difficulty of describing the complex dynamics of CMEs during the interplanetary phase. To date, a wide variety of approaches has been developed to investigate CME evolution. These range from rudimentary empirical models \citep{gopalswamy2000interplanetary} and semiempirical models \citep{mostl2018forward}, to full magnetohydrodynamic simulations (cf. \citealp{pomoell2018euhforia}), to simplified kinematic descriptions such as the drag-based model (DBM; \citealp{vrvsnak2002influence,vrvsnak2013propagation}), and more recently to data-driven machine learning and deep learning (ML/DL) algorithms (e.g., \citealp{alobaid2023estimating,chierichini2024cme,wang2019cme,yang2023prediction}; see also \citealp{camporeale2019challenge} for a detailed discussion of their limitations). Each of these models has its own strengths and limitations. As in many other areas of applied science, recent developments have naturally led to the integration of ML techniques with deterministic models under the name of physics-informed ML, in an effort to strike a balance between accuracy and computational efficiency, particularly for operational nowcasting and forecasting purposes. A successful implementation of this strategy is presented in \citet{guastavino2023physics}, where the DBM was employed as the physical backbone of the method, achieving a reduction of approximately 1-2 hours in the mean absolute error of arrival time predictions compared to previous benchmarks (c.f. \citealp{riley2018forecasting}). 

The DBM-based artificial intelligence (AI) approach clearly inherits the intrinsic limitations of the DBM itself. In particular, it cannot account for CME events that may have experienced during their propagation additional accelerations (or decelerations) beyond solely the drag (e.g., residual accelerations due to internal magnetic reconfiguration; see \citealp{manchester2017physical} for a thorough illustration of the processes responsible for this effect). Such cases are not uncommon and are not adequately captured by the DBM \citep{regnault2024discrepancies}. 

Building on the work of \citet{guastavino2023physics}, the present study aims to improve predictive performance by testing a loss function capable of retaining in the training process CME events that would normally be discarded when using the classical DBM. A natural choice consists of incorporating the extended drag-based model (EDBM), recently proposed by \citet{rossi2025extended}, which is dynamically richer and more robust, making it well suited for this purpose.

Since the available LASCO observational dataset is dominated by CME events that either asymptotically approach the background solar wind (DBM-compatible events) or decelerate below it (non-DBM-compatible events), in this paper, we leveraged the EDBM to design a neural network capable of processing CME events belonging to the latter category. The core methodological component is a neural network that predicts the travel time of CMEs to 1 AU, and which is trained using a physics-constrained loss function originally developed by \citet{guastavino2023physics}, in which the DBM-based physics term is replaced by the corresponding EDBM-based counterpart. Specifically, we first obtained an approximation of the actual (unknown) non-drag deceleration using the EDBM solution applied to the training set. The neural network was then trained to estimate the CME travel time to 1 AU, taking as input only the observed CME parameters, i.e., the CME initial speed, mass, and impact area, and solar wind parameters, i.e., the solar wind speed and density. The previously computed deceleration was treated as an auxiliary parameter, used solely during the optimization of the loss function. This design ensures that, once trained, the model can infer travel times for unseen events without requiring a priori knowledge of any additional, possibly present, underlying negative acceleration. To address data scarcity, we applied data augmentation techniques to expand both the training and validation sets. To evaluate the robustness of our results, we performed 25 realizations of training, validation, and test splits using a uniform sampling strategy that preserves the distributions of solar wind speed, distinguishing between ``fast'' ($\ge500$ km/s) and ``slow'' ($<500$ km/s) solar wind (while remaining within the dynamical regime where the initial CME speed exceeds the solar wind speed). Furthermore, we assessed the robustness of our approach under parameter uncertainty by performing a sensitivity analysis on the drag coefficient, the most critical yet unobservable parameter in the EDBM model (see Sect.~\ref{sec:edbm}). We find that the EDBM-guided AI strategy maintains comparable or superior performance to purely data-driven or physics-based approaches and achieves results similar to those of the DBM-guided AI model, with only a slight decrease in accuracy.

The EDBM defines multiple propagation regimes (including DBM-compatible ones), depending on the initial and final CME front speed relative to the solar wind (see Sect.~\ref{sec:edbm}). Although the methodology presented here is restricted to one speed regime (consisting of non-DBM-compatible CME events), it can readily be applied to all regimes. For operational forecasting applications, incorporating all EDBM regimes is necessary, since even regimes currently sparsely populated with data may correspond to future CME events. Nevertheless, each EDBM regime entails a specific deterministic formulation; therefore, a classification algorithm is required to assign each new CME event to the appropriate EDBM class. To this end, we propose a multinomial logistic-regression classifier fed with CME measured input features. This model could ultimately be integrated as the initial stage of a potentially operational CME travel-time prediction pipeline.

The paper is structured as follows. Sect.~\ref{sec:edbm} briefly reviews the dynamics of the EDBM and related speed regimes. Sect.~\ref{sec:ML} outlines the ML workflow for travel-time prediction from a conceptual perspective, detailing the neural network architecture and optimization procedure. Sect.~\ref{sec:exper} presents the application of the above methodology to a subset of LASCO observations that fall within the speed regime of interest. Sect.~\ref{sec:multiclass} reports the results of the multiclass classification algorithm applied to the full set of LASCO observations, which include both DBM-compatible and non-DBM-compatible events spanning different speed regimes. Concluding remarks and future developments are offered in Sect.~\ref{sec:concl}.

\section{The extended drag-based model}
\label{sec:edbm}
Born as a natural extension of the DBM developed by \citet{cargill2004aerodynamic, vrvsnak2013propagation}, the EDBM \citep{rossi2025extended} has been applied to CME heliospheric evolution to overcome the dynamical limitations of the original DBM. Despite the increased complexity, the model still maintains a closed-form solution. This extension introduces a constant additional acceleration term $a$, which can be either positive or negative, accounting for forces not captured by the aerodynamic drag term alone (e.g., magneto-gravitational forces):
\begin{equation}
    \ddot{r} = -\gamma |\dot{r}-w|(\dot{r}-w)+a\;,
    \label{eq:a}
\end{equation}
where $r=r(t)$ is the CME radial position at time $t$, $w = w(r,t)$ is the solar wind speed and $\gamma$ is the drag parameter. For $a=0$, Eq.~\eqref{eq:a} corresponds to the standard DBM. As in the case of the DBM, a unique analytical solution to the Cauchy problem defined by Eq.~\eqref{eq:a} with initial conditions $r(0)=r_0$, $\dot{r}(0)=v_0$ exists, provided that the quantities in the drag contribution $-\gamma |\dot{r}-w|(\dot{r}-w)$ are assumed to be constant. These are the drag parameter $\gamma$, which can be expressed as $\gamma=CA\rho/m$, and the solar wind speed $w$. In the expression of $\gamma$, $C$ is the drag coefficient, $A$ is the cross-sectional (or impact) area of the CME, $\rho$ is the ambient solar wind density, and $m$ is the mass of the CME. 

The sign of $a$ together with the value of the initial speed $v_0$ uniquely determine the dynamical regime of the CME, as well as the corresponding form of the solution to Eq.~\eqref{eq:a} and its time domain of validity (cf. Table~\ref{tab:CMEcases} and \citealp{rossi2025extended}). In particular, the classical DBM requires that the propagation speed $v(t)$ of each CME event satisfies
\begin{equation}
\label{eqn:condDBM}
v_0 <v(t) < w \quad \text{or} \quad v_0 > v(t) > w\quad \text{or}\quad v(t)\equiv v_0=w\;,
\end{equation}
for every time $t>0$. Instead, the EDBM relaxes this requirement in essentially two ways: either by allowing the CME speed to deviate from the asymptotic wind speed $w$, i.e., to exceed or fall below it, or by allowing motion against it over time. Both of these dynamical regimes are incompatible with the DBM. For this reason, we classify CME dynamical regimes according to Table~\ref{tab:CMEcases}, distinguishing between events that satisfy the condition \eqref{eqn:condDBM} and those that do not.
\begin{table*}
\caption{Categorization of CME propagation scenarios according to the EDBM, based on the CME initial speed $v_0$, final speed $v$, and solar wind speed $w$.}
\renewcommand{\arraystretch}{3.7}
\centering
\begin{tabular}{cccccp{3.5cm}}
\toprule\toprule
\addlinespace[-8pt]
\textbf{Case} & \textbf{Speed condition} & \textbf{Model} & \textbf{Sign of }$a$ & \textbf{Time domain for }$r(t)$ & \textbf{CME description}\\
\midrule
Sub-wind ($\nearrow$) & $v_0, v \leq w,\ v_0 \le v$ & EDBM/DBM & $+$ & $\displaystyle\left[0,\frac{\arctan\left(\sqrt{\gamma/a}(w-v_0)\right)}{\sqrt{a\gamma}}\right]$ & \parbox[c][2em][c]{3.4cm}{Slower than the solar wind, accelerating toward it}\\
Sub-wind ($\searrow$) & $v_0, v \leq w,\ v_0 > v$ & EDBM & $-$ & $\displaystyle[0,+\infty)$ & \parbox[c][2em][c]{3.5cm}{Slower than the solar wind, further decelerating}\\
Super-wind ($\nearrow$) & $v_0, v \geq w,\ v_0 \le v$ & EDBM & $+$ & $\displaystyle[0,+\infty)$ & \parbox[c][2em][c]{3.4cm}{Faster than the solar wind, further accelerating}\\
Super-wind ($\searrow$) & $v_0, v \geq w,\ v_0 > v$ & EDBM/DBM & $-$ & $\displaystyle\left[0,\frac{\arctan\left(\sqrt{-\gamma/a}(v_0-w)\right)}{\sqrt{-a\gamma}}\right]$ & \parbox[c][2em][c]{3.4cm}{Faster than the solar wind, decelerating toward it} \\
Cross-wind ($\nearrow$) & $v_0 < w < v$ & EDBM & $+$ & $\displaystyle\left(\frac{\arctan\left(\sqrt{\gamma/a}(w-v_0)\right)}{\sqrt{a\gamma}},+\infty\right)$ & \parbox[c][2em][c]{3.4cm}{Initially slower than the solar wind, later exceeding it} \\
Cross-wind ($\searrow$) & $v_0 > w > v$ & EDBM & $-$ & $\displaystyle\left(\frac{\arctan\left(\sqrt{-\gamma/a}(v_0-w)\right)}{\sqrt{-a\gamma}},+\infty\right)$ & \parbox[c][2em][c]{3.4cm}{Initially faster than the solar wind, later falling below it}\\
\addlinespace[12pt]
\bottomrule
\end{tabular}
\tablefoot{Cases Sub-wind$(\nearrow)$ and Super-wind$(\searrow)$ include DBM-modelable events ($a=0$), as they fulfill the condition \eqref{eqn:condDBM}, whereas the remaining cases require the extended formulation exclusively.}
\label{tab:CMEcases}
\end{table*}

We conclude this section by recalling the expression of the solution $r(t)$ in the case where $v_0>w$ and the CME decelerates below $w$, a situation we refer to as the Cross-wind $(\searrow)$ case, which is useful in the next section:
\begin{align}
\label{eqn:rcrosswinddown}
    &r_{\text{Cross-wind ($\searrow$)}}(t)=r_0+\left(w-\sqrt{-\frac{a}{\gamma}}\right)t \nonumber \\
    &-\frac{1}{\gamma}\Bigg(\ln\left(\frac12\sqrt{\frac{a}{a-\gamma(v_0-w)^2}}\left(e^{-2\left(\sqrt{-a\gamma}t-\arctan\left(\sqrt{-\gamma/a}(v_0-w)\right)\right)}+1\right)\right) \nonumber \\
    &\hspace{4cm}-\arctan\left(\sqrt{-\frac{\gamma}{a}}(v_0-w)\right)\Bigg)\;.
\end{align}
Note that events that satisfy Eq.~\eqref{eqn:condDBM} can also be modeled within the EDBM framework, provided that $|a|$ is typically small compared to the drag acceleration, although not necessarily. In fact, by the continuous dependence on the parameter $a$ of Eq.~\eqref{eq:a}, the limit $a\to0$ in $r(t)$ correctly recovers the DBM solution (cf. \cite{vrvsnak2013propagation}) for the DBM-compatible cases, here referred to as Sub-wind $(\nearrow)$ and Super-wind $(\searrow)$ (see Table~\ref{tab:CMEcases}).

\section{Machine learning workflow for travel-time prediction}
\label{sec:ML}
In the following, we propose a two-stage procedure that integrates physics-based modeling with machine learning for operational prediction of CME travel times applied to CME events belonging to the Cross-wind $(\searrow)$ case. 

The input feature vector used in the two stages consists of direct and indirect measurements; in our case, it is given by $\boldsymbol{x}=(v_0,m,A,\rho,w)$, whilst $r_0$ and $C$ are treated as event-independent fixed parameters. This assumption is motivated by the fact that, while a reference measure for $r_0$ is generally available (typically the eruption distance), the dimensionless coefficient $C$ is neither observable nor directly computable from data. Magnetohydrodynamic simulations and probabilistic approaches indicate that $C$ varies slowly between the Sun and 1 AU, and ranges from 1 to 100 for most CMEs \citep{cargill2004aerodynamic,napoletano2018probabilistic}, leading to an average value of $\gamma$ of about $1\times10^{-7}$ km$^{-1}$ \citep{napoletano2022parameter}. However, this parameter is crucial for DBM-type approaches \citep{chierichini2024bayesian}, including both fully physics-based methods and physics-driven AI models \citep{guastavino2023physics}, where it plays an important role in improving travel-time predictions. For this reason, in Sect.~\ref{subsec:C} we analyze how EDBM-based AI responds to variability in $C$ using experimental data.

\subsection{Stage (i): non-drag acceleration estimation}
\label{subsec:stagei}
The first stage of the procedure involves computing the acceleration parameter $a$, which represents an estimate of the actual additional acceleration experienced by the CME. This quantity is computed by numerically solving for $a$ the equation 
\begin{equation}
\label{eqn:requal1}
r_{\text{Cross-wind ($\searrow$)}}(t;a,\boldsymbol{x})=1~\mathrm{AU}\;,
\end{equation}
where the left-hand side is given by Eq.~\eqref{eqn:rcrosswinddown}. This computation is carried out exclusively for training and validation samples, as it involves the computation of the loss function, which is optimized and monitored throughout the training and validation phases (see Sect.~\ref{subsec:stageii}). During these phases, in addition to the input features $\boldsymbol{x}$, the observed travel time $t$ is known.

Events for which no valid solution exists, either due to non-convergence of the numerical solver (e.g., standard root-finding Newton-Raphson methods; cf. \citealp{suli2003introduction}) or incompatibility with the corresponding time domain in Table~\ref{tab:CMEcases}, are excluded from the training set.

\subsection{Stage (ii): EDBM-informed ML for travel-time prediction}
\label{subsec:stageii}
The above resulting acceleration $a$ is now incorporated as an auxiliary quantity in the second stage of the procedure, which focuses on predicting the CME travel time using a feedforward neural network (e.g., \citealp{goodfellow-et-al-2016}, Chapter 6). To ensure physically consistent predictions, we adopt a physics-informed learning approach, where physical information is embedded directly into the loss function. The network is trained using a composite loss function, consisting of two weighted contributions, namely a data-driven term and a physical model-driven term \citep{guastavino2023physics}:
\begin{align}
\label{eqn:loss}
\mathcal{L}(\hat{t}(\boldsymbol{x});t,a,\boldsymbol{x}) = &\lambda_1 k_1(\hat{t}(\boldsymbol{x})-t)^2 \nonumber\\
&+\lambda_2k_2(r_\text{Cross-wind ($\searrow$)}(\hat{t}(\boldsymbol{x});a,\boldsymbol{x}) - 1~\text{AU})^2\;,
\end{align}
where the data-driven term quantifies the quadratic deviation between the observed travel time $t$ and the predicted output travel time $\hat{t}$ returned by the neural network, while the model-driven term quantifies the quadratic deviation between the distance traveled according to the EDBM and 1 AU. We set weights $\lambda_1 = \lambda_2 = 1/2$ to assign equal importance to the empirical and physical components of the loss, while $k_1$ and $k_2$ are normalization constants that ensure that the two components operate on comparable dimensionless scales.

The architecture of the neural network $\hat{t}(\boldsymbol{x})$, described in Table~\ref{tab:architecture}, was designed to extract hierarchical representations, i.e., lower layers capture simple patterns among the input features (e.g., linear relationships), while deeper layers combine them into more complex relationships. The network consists of eight hidden layers with progressively decreasing numbers of neurons (units). To mitigate overfitting, regularization techniques, such as dropout layers, were also implemented. In our architecture, dropout layers are inserted after the first two hidden layers, respectively, which randomly deactivate 40\% of the neurons in the preceding layer during each training epoch. We used the rectified linear unit activation function (ReLU) $f(x)=\max\{0,x\}$ to introduce nonlinearity throughout the network. The output layer employs a Softplus activation function $f(x)=\log(1+\exp(x))$ to ensure that predicted travel times are strictly positive. 
\begin{table}
\caption{Details of the neural network architecture for travel-time prediction in Stage (ii) (see text).}
\centering
\begin{tabular}{lll}
\toprule\toprule
\textbf{Layer} & \textbf{Units} & \textbf{Activation} \\
\midrule
Input      & 5   & ---       \\
Dense 1    & 200 & ReLU      \\
Dropout    & --- & 0.4       \\
Dense 2    & 150 & ReLU      \\
Dropout    & --- & 0.4       \\
Dense 3    & 100 & ReLU      \\
Dense 4    & 75  & ReLU      \\
Dense 5    & 50  & ReLU      \\
Dense 6    & 30  & ReLU      \\
Dense 7    & 25  & ReLU      \\
Dense 8    & 10  & ReLU      \\
Output     & 1   & Softplus  \\
\bottomrule
\end{tabular}
\label{tab:architecture}
\end{table}

The network was trained using the Adam optimizer with a learning rate of $10^{-3}$ and Glorot uniform weight initialization \citep{glorot2010understanding,kingma2015adam}. Training runs for a maximum of 1000 epochs, with early stopping based on validation loss and a patience of 100 epochs to prevent overfitting \citep{caruana2001overfitting}.
The implementation was carried out using Python TensorFlow software library \citep{abadi2016tensorflow}.

\section{Application to EDBM-modelable real observations}
\label{sec:exper}
We used the dataset developed by \citet{napoletano2022parameter}, which includes CMEs selected by combining information from the Richardson and Cane CME list \citep{richardson2010near} and the SOHO-LASCO catalog \citep{yashiro2004catalog}. This dataset provides a variety of parameters necessary for application in DBM-type models. Some quantities are directly extracted from the source lists, while others are derived as part of the results presented in \cite{napoletano2022parameter}. The dataset includes, among other parameters, the CME arrival speed $v$, initial speed $v_0$, travel time $t$, angular width (from which the impact area $A$ can be computed), and mass $m$. As estimates for the solar wind speed $w$ and density $\rho$, we adopted the values used in \citet{guastavino2023physics}, which are derived from CELIAS data. We used these values of $w$, along with the CME initial and arrival speeds, to classify CME events into different EDBM regimes (Table~\ref{tab:CMEcases}). 

The complete dataset contains a total of 213 CME events in the temporal range between 1997 and 2018, but we restricted our analysis to the 160 events for which mass information is available, as reported in \citet{guastavino2023physics}. All CME and solar wind parameters were treated as fixed over the CME propagation, assuming spatial and temporal uniformity. The CME onset height (i.e., eruption height) was fixed at $r_0 = 20R_\odot$, where $R_\odot=695\,700$ km is the solar radius, following standard assumptions in drag-based modeling \citep{vrvsnak2014heliospheric}. For convenience, the CME eruption time was set as the starting time ($t=0$), as in Sect.~\ref{sec:edbm}.

To ensure physically plausible values for the drag coefficient $\gamma$, the empirical constant $C$ used in the computation of $\gamma$ was set to 100, based on physical assumptions and a series of calibration tests (details are provided in Sect.~\ref{subsec:C}). This choice yields $\gamma$ values on the order of $10^{-7}$--$10^{-8}$ km$^{-1}$, with a mean of $\gamma=1.6\times10^{-7}$ km$^{-1}$ (Table~\ref{tab:travel_time_C}), consistent with values previously reported in the literature \citep{vcalogovic2025constraints,napoletano2022parameter}.

Each CME was categorized into one of the propagation regimes defined in Sect.~\ref{sec:edbm}. The number of events per case is reported in Table~\ref{table:numberofevents}. Notably, the Cross-wind $(\searrow)$ events (a class that necessarily requires the EDBM formulation) account for roughly 25\% of the dataset, underscoring the need for a more general modeling framework than the classical DBM. In the remainder of the section, we apply the methodology described in Sect.~\ref{sec:ML} to these 41 events, leaving aside the remaining events of the dataset. The complete 160-event dataset was used for the speed-regime multiclass classification described in Sect.~\ref{sec:multiclass}.
\begin{table}
\caption{Number of CME events per EDBM propagation case.}
\centering
\begin{tabular}{cc}
\toprule\toprule
\textbf{Case} & \textbf{Number of Events} \\
\midrule
Sub-wind ($\nearrow$) & 9 \\
Sub-wind ($\searrow$) & 6 \\
Super-wind ($\nearrow$) & 0 \\
Super-wind ($\searrow$) & 96 \\
Cross-wind ($\nearrow$) & 8 \\
Cross-wind ($\searrow$)  & 41 \\
\bottomrule
\end{tabular}
\label{table:numberofevents}
\end{table}

\subsection{Data augmentation and dataset stratification}
\label{subsec:aug}
Given the limited number of available Cross-wind $(\searrow)$ events for proper ML training, we employed data augmentation to enhance model robustness and reduce overfitting. During training and validation, each input feature was randomly perturbed by a value within $\pm5\%$ to $\pm10\%$ of its original magnitude. For each real data point, 100 augmented data samples were generated. A consistency check was then performed to ensure that the augmented data met the required speed condition for the Cross-wind $(\searrow)$ case ($v_0>v>w$) and that a nonzero extra deceleration ($a<0$) could be found. This controlled noise injection introduces variability while preserving the underlying physical plausibility of the samples. Empirical testing showed that this setup provided the best trade-off between generalization and model stability.

The dataset was also stratified according to solar wind speed to ensure balanced sampling. Two probability density functions are commonly considered, corresponding to two solar wind ensembles originating from different regions of the Sun (e.g., \citealp{schwenn2006space}): the slow solar wind ($w<500$ km/s) and the fast solar wind ($w\geq500$ km/s). We adopted the same distinction to avoid training bias linked to propagation environment: CME events were labeled as slow-wind (i.e., CMEs traveling with slow solar wind) or fast-wind (i.e., CMEs traveling with fast solar wind).

\subsection{Travel-time prediction for Cross-wind $(\searrow)$ CME events}
\label{subsec:TT}
To assess the predictive performance of our methodology on the Cross-wind $(\searrow)$ events, we carried out 25 independent experimental runs following the methodology outlined in Sect.~\ref{sec:ML}. Due to the scarcity of real-world observations for this class, training and validation were performed on a combined dataset that included both real and augmented samples, according to Sect.~\ref{subsec:aug}. However, to provide a more accurate assessment of model generalization, performance metrics are reported separately for real and augmented sets.

We adopted three complementary evaluation metrics for each realization (i.e., for each experimental run): mean absolute error (MAE), median absolute error (MedAE), and relative error. This choice accounts for the limited sample size and improves robustness to potential outliers, with the MedAE being particularly informative in small-data regimes.

\begin{figure*}
 \centering
 \includegraphics[width=0.33\linewidth]{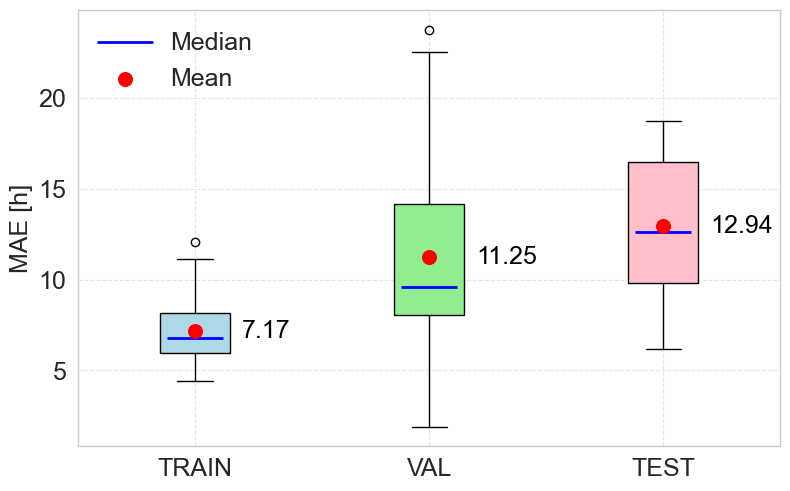}
 \includegraphics[width=0.33\linewidth]{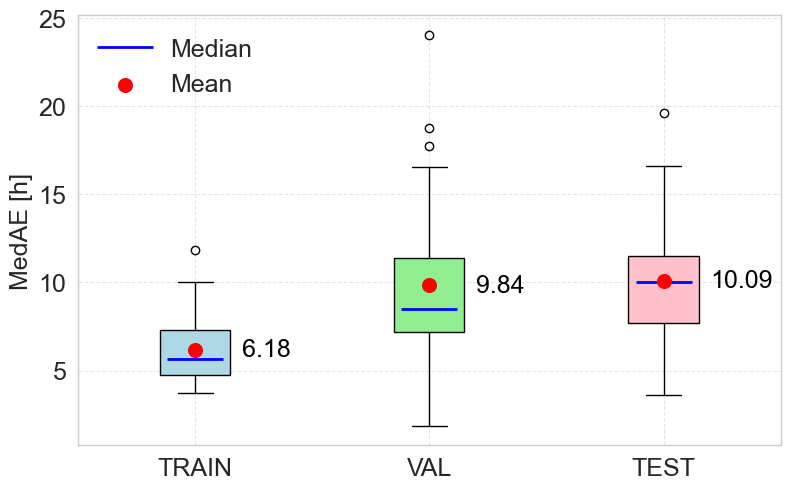}
 \includegraphics[width=0.33\linewidth]{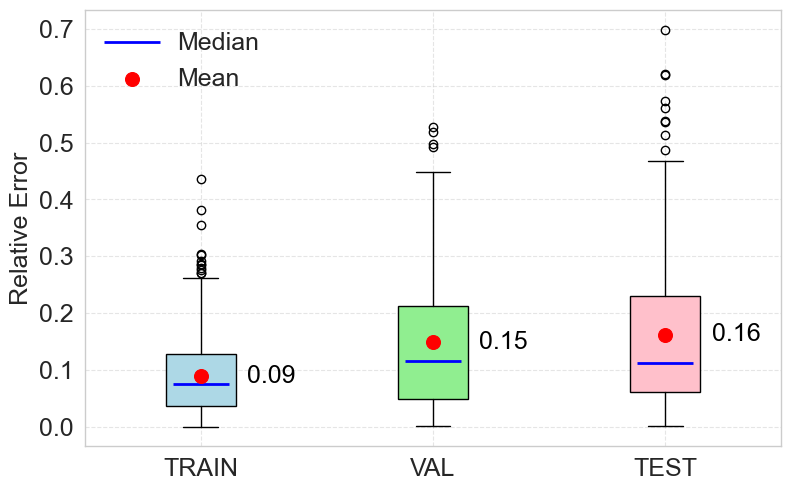}
 \includegraphics[width=0.33\linewidth]{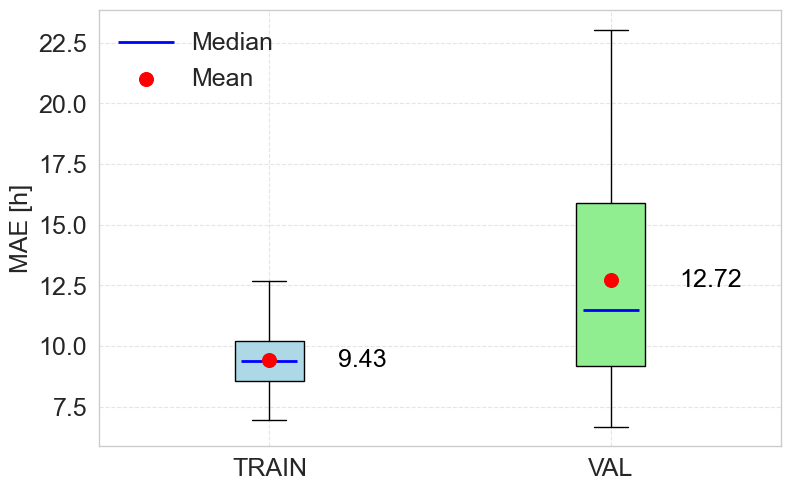}
 \includegraphics[width=0.33\linewidth]{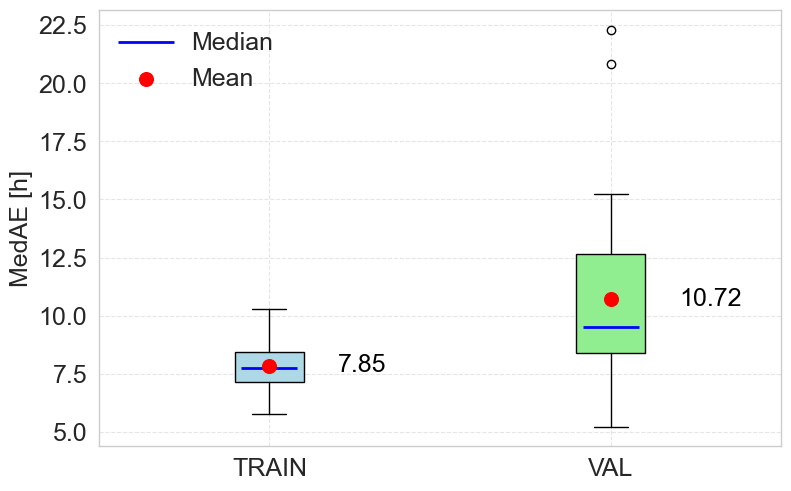}
 \includegraphics[width=0.33\linewidth]{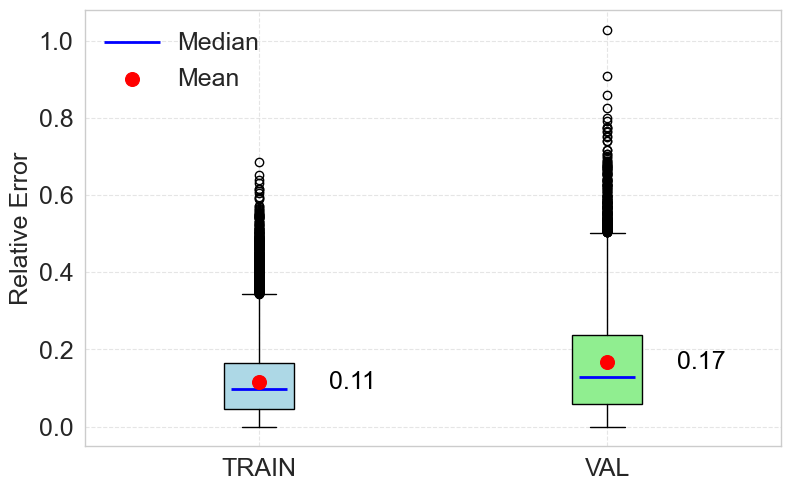}
 \caption{Prediction errors on Cross-wind ($\searrow$) events across 25 runs for training (light blue box), validation (green box), and test (red box) sets. Top panels: real data. Bottom panels: augmented data. From left to right: mean absolute error (MAE), median absolute error (MedAE), and relative error. Numbers in the plots indicate mean values over the 25 realizations.}
\label{fig:real_vs_aug}
\end{figure*}

The box plots in Fig.~\ref{fig:real_vs_aug} show the error distributions we obtained across all runs, with real test-set data in the top panels and augmented training/validation data in the bottom panels. For real data, despite the limited number of available samples, the results remain consistent for all runs. The MedAE often falls below the MAE, highlighting the model’s resilience to extreme deviations; in other words, extreme outliers do not strongly influence the model. Remarkably, relative errors remain consistently on average within the 16\%, in line with performance benchmarks reported in recent CME travel-time prediction studies \citep{guastavino2023physics, wang2019cme}.

Augmented data exhibits similar behavior with slightly reduced variance, reflecting the regularizing effect of the augmented training data. The narrower spread of errors indicates improved stability without compromising accuracy. On average, the model achieves travel-time predictions within about 13 hours of the observed arrivals, demonstrating competitive accuracy given the natural intrinsic variability of CME propagation.

\begin{figure}
    \centering
    \begin{subfigure}{0.242\textwidth}
    \centering
    \includegraphics[width=\textwidth]{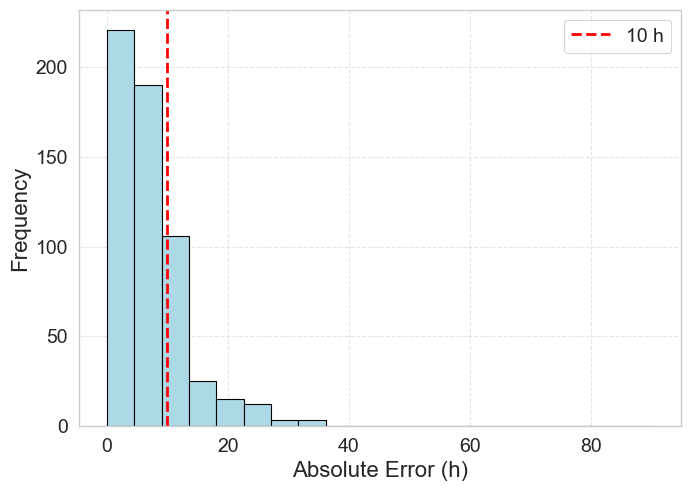}
    \subcaption{Training (real)}
    \end{subfigure}
    \begin{subfigure}{0.242\textwidth}
    \centering
    \includegraphics[width=\textwidth]{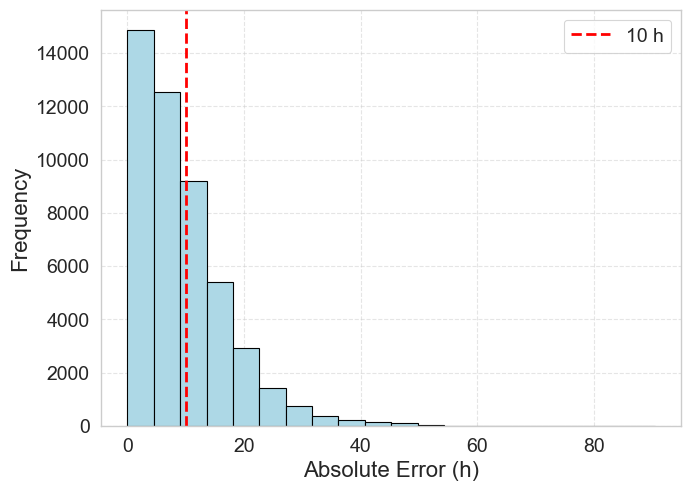}
    \subcaption{Training (augmented)}
    \end{subfigure}
    \vspace{-1mm}
    
    \begin{subfigure}{0.242\textwidth}
    \centering
    \includegraphics[width=\textwidth]{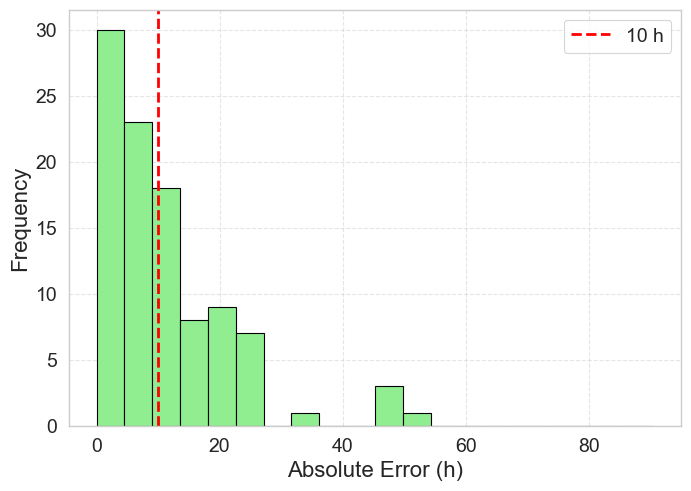}
    \subcaption{Validation (real)}
    \end{subfigure}
    \begin{subfigure}{0.242\textwidth}
    \centering
    \includegraphics[width=\textwidth]{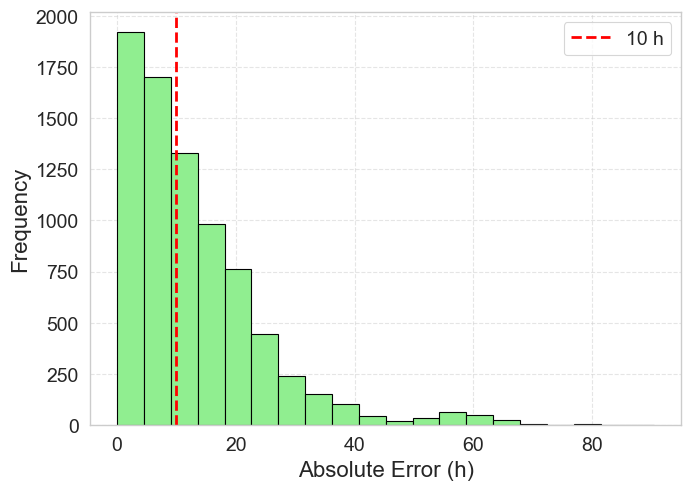}
    \caption{Validation (augmented)}
    \end{subfigure}
    \vspace{-1mm}
    
    \begin{subfigure}{0.242\textwidth}
    \centering
    \includegraphics[width=\textwidth]{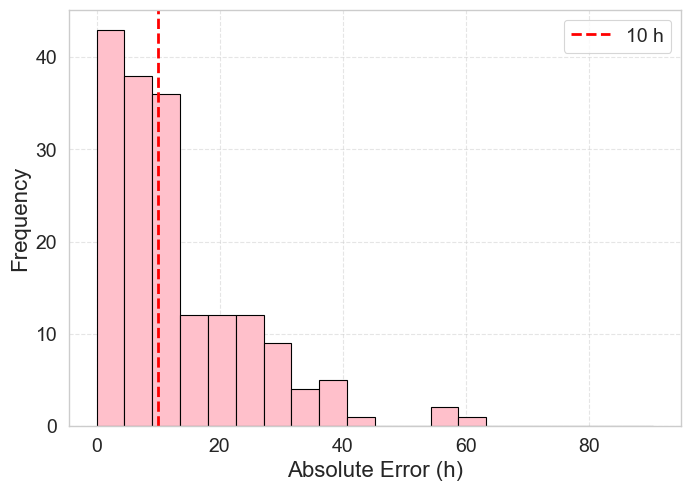}
    \subcaption{Test}
    \end{subfigure}
    \caption{Absolute-error distributions of Cross-Wind $(\searrow)$ events across 25 runs. Plots (a), (c), and (e): real events. Plots (b) and (d): augmented events. The red dashed line marks the 10-hour threshold.}
    \label{fig:hist}
\end{figure}

To further illustrate the distributions of prediction errors, Fig.~\ref{fig:hist} compares the absolute-error histograms for real and augmented events across all runs. The model’s performance improves as the error threshold increases. For $\mathrm{MAE}<10$ h, $50$–$75\%$ of predictions fall within range, with $74.6\%$ for real training data (plot (a)), $53\%$ for real validation data (plot (c)), and $50.3\%$ for (real) test data (plot (e)). At the MAE threshold of 12.94 h (the average MAE obtained for real data shown in the top-left panel of Fig.~\ref{fig:real_vs_aug}), $65$–$86\%$ of predictions are accurate, while for $\mathrm{MAE}<15$ h, $67$–$91\%$ of predictions lie within the threshold. The augmented samples (plots (b) and (d)) show a tighter clustering around the median. This behavior is expected, as augmented data resemble more closely the training distribution, whereas real events provide a more challenging test of the model’s generalization.

\subsection{Sensitivity to the drag coefficient}
\label{subsec:C}
In order to address the role of the empirical constant $C$ in the EDBM-based AI model, we performed a sensitivity analysis by varying its value within the range $80$--$160$, with spacing of 10. This interval was selected based on both physical and experimental considerations. From a physical viewpoint, assuming comparable CME and solar wind densities, $C$ is expected to increase to values much larger than unity and remain approximately constant with radial distance between the Sun and 1 AU \citep{cargill2004aerodynamic}. Additionally, since we consider CMEs initially faster than the solar wind and subsequently decelerated, these events correspond to larger values of $\gamma$, as they are expected to experience stronger drag \citep{napoletano2022parameter,vrvsnak2008dynamics,vrvsnak2013propagation}, consistent with the upper part of the range $0.1$--$3\times 10^{-7}$ km$^{-1}$ \citep{chierichini2024bayesian}). The selected range of $C$ therefore allows us to investigate model performance for mean values of the drag parameter lying approximately within this upper interval (see Table~\ref{tab:travel_time_C}). Numerically, for each tested value of $C$, the model was retrained and evaluated using the same methodology, configuration, and dataset described in Sect.~\ref{sec:ML} and in the present section. As shown in Table~\ref{tab:travel_time_C}, the average MAE on the test set exhibits a shallow minimum around $C=100$, indicating that this value represents an optimal compromise between physical plausibility and predictive performance.
\begin{table*}
\caption{Mean absolute error (MAE) and median absolute error (MedAE) for travel-time predictions on real Cross-wind $(\searrow)$ data for different values of $C$.}
\centering
\begin{tabular}{ccccccccc}
\toprule\toprule
\multirow{2}{*}{$C$} &
\multicolumn{2}{c}{\textbf{Training [h]}} &
\multicolumn{2}{c}{\textbf{Validation [h]}} &
\multicolumn{2}{c}{\textbf{Test [h]}} &
\multirow{2}{*}{\textbf{Mean} $\gamma$ \textbf{[km$^{-1}$]}} &
\multirow{2}{*}{\textbf{Valid events}} \\
 & MAE & MedAE & MAE & MedAE & MAE & MedAE & & \\
\midrule
80 & 6.89 & 5.88 & 10.29 & 8.73 & 13.67 & 10.42 & $1.3\times10^{-7}$ & 34\\
90 & 7.41 & 6.21 & 11.11 & 9.45 & 14.05 & 11.65 & $1.4\times10^{-7}$ & 35\\
\textbf{100} &  \textbf{7.17}  &  \textbf{6.18}  &  \textbf{11.25}  &  \textbf{9.84}  &  \textbf{12.94}  &  \textbf{10.09}  & $\boldsymbol{1.6\times10^{-7}}$ & \textbf{34} \\
110 & 6.73 & 5.49 & 10.77 & 9.36 & 13.94 & 11.42 & $1.7\times10^{-7}$ & 35\\
120 & 7.18 & 6.24 & 10.68 & 9.31 & 13.72 & 11.07 & $1.9\times10^{-7}$ & 35\\
130 & 7.15 & 5.86 & 10.92 & 9.16 & 13.92 & 10.96 & $2.2\times10^{-7}$ & 35\\
140 & 7.47 & 6.55 & 11.18 & 9.24 & 13.47 & 10.57 & $2.2\times10^{-7}$ & 35 \\
150  &  7.00  &  5.83  &  11.03  &  8.99  &  13.38  &  10.83  & $2.3\times10^{-7}$ & 35 \\
160 & 7.28 & 6.06 & 10.67 & 9.28 & 13.66 & 11.39 & $2.5\times10^{-7}$ & 35\\
\bottomrule
\end{tabular}
\tablefoot{MAE and MedAE for training, validation and test sets were averaged over 25 realizations. Mean values of $\gamma$ were computed over the reported valid events (see text). The bold row corresponds to the mean metrics in the top-left and top-middle panels of Fig.~\ref{fig:real_vs_aug}.}
\label{tab:travel_time_C}
\end{table*}

Referring to the same table, the dependence of the MAE and MedAE on $C$ indicates that the model is moderately sensitive to the calibration of this empirical constant, yet remains sufficiently stable within the explored range. Values of the drag coefficient outside this range result in either underfitting, with increased training and test errors, or physically inconsistent behavior for smaller $C$. For larger $C$, signs of overfitting appear, where the training MAE continues to decrease without a corresponding improvement in test performance. Notably, valid real CME events, namely those for which a meaningful deceleration $a$ was computed according to Eq.~\eqref{eqn:requal1} and Sect.~\ref{subsec:stagei}, appear unaffected by variations in $C$.

\section{Multiclass classification of speed regimes}
\label{sec:multiclass}
In Sect.~\ref{sec:exper}, we focused exclusively on the Cross-wind $(\searrow)$ propagation class, as the remaining EDBM-specific classes were underrepresented in the dataset (Table~\ref{table:numberofevents}). However, the AI approach developed for travel-time prediction can be readily extended to other propagation regimes, including those compatible with the standard DBM. This can be achieved by separating parallel neural networks, which have to be trained on the corresponding CME event subsets and by employing the associated analytical solution of the EDBM into the loss function, similarly to Eq.~\eqref{eqn:loss}. Hence, during inference, each new input CME event must be appropriately sorted into one of the speed regimes, despite the final speed $v$ being unknown.

In the present section, we construct a classification module that performs this task. We adopt a standard multinomial logistic regression classifier (e.g., \cite{kleinbaum2002logistic}), considering a six-class classification setup that distinguishes the six propagation regimes defined within the EDBM framework in Table~\ref{tab:CMEcases}. The classifier was implemented using the Python Scikit-Learn library \citep{scikit-learn} and can be used operationally in the event a new CME occurs.

Conceptually, given the set of propagation regimes 
\begin{multline*}
\mathscr{C}=\{\text{Sub-wind ($\nearrow$)},\text{Sub-wind ($\searrow$)},\text{Super-wind ($\nearrow$)},\\
\text{Super-wind ($\searrow$)},\text{Cross-wind ($\nearrow$)},\text{Cross-wind ($\searrow$)}\}\;,
\end{multline*}
we denote by $\{(\boldsymbol{x}_i, y_i)\}_{i=1}^n$ the set of $n$ samples, where $\boldsymbol{x}_i \in \mathbb{R}^d$, $d=5$, denotes the usual input feature vector (see Sect.~\ref{sec:ML}) associated with the CME event $i$; $y_i \in \mathscr{C}$ are the corresponding class labels, where the true class of the CME is determined from the observed final speed $v$ and initial speed $v_0$. Let $\boldsymbol{W} \in \mathbb{R}^{|\mathscr{C}| \times d}$ and $\boldsymbol{b} \in \mathbb{R}^{|\mathscr{C}|}$ denote the model parameters, i.e., the weights and biases, respectively, where $|\mathscr{C}| = 6$ is the number of classes. We define the linear score for class $c \in \mathscr{C}$ and input sample $i$ as
\begin{equation}
\label{eqn:score}
    z_{i,c} = \boldsymbol{W}_c\cdot \boldsymbol{x}_i + b_c\;,
\end{equation}
where $\boldsymbol{W}_c\in\mathbb{R}^d$ is the row vector of weights for class $c$ and $b_c$ is the corresponding scalar bias. The multinomial logistic regression model assigns to the sample with feature vector $\boldsymbol{x}_i$ the probability of belonging to class $c\in \mathscr{C}$ as
\begin{equation}
\label{eqn:prob}
    p_{i,c}=\frac{\exp(z_{i,c})}{\sum_{j=1}^{|\mathscr{C}|} \exp(z_{i,j})}\;.
\end{equation}
The model parameters are then estimated by minimizing the multinomial negative 
log-likelihood, defined as
\begin{equation}
\label{eqn:loglike}
    \mathscr{L}(\boldsymbol{W},\boldsymbol{b})
    = -\frac{1}{n} \sum_{i=1}^{n} \log p_{i,y_i}\;,
\end{equation}
and the predicted class $\hat{c}_i$ for sample $i$ is given by the class $c$ that maximizes $p_{i,c}$.

During inference, only the input $\boldsymbol{x}$ is required. In addition, to address class imbalance, the classifier is trained using a stratified sampling strategy analogous to the one introduced in \cite{2023FrASS...939805G} for solar flare forecasting. In this approach, the training set is constructed so that each class contributes samples in proportion to its true frequency, rather than allowing the majority class to dominate. This ensures that all propagation regimes, including the less frequent ones, are adequately represented during training, reducing bias and improving the classifier’s ability to identify rare but operationally important events. 

Model performance is evaluated using the accuracy metric and the true skill statistic (TSS) for each predicted output class $\hat{c}$, the latter being a metric that is well suited for imbalanced classification problems. The Accuracy measures the fraction of CME events whose propagation regime is correctly classified by the model. The TSS is defined as
\begin{equation}
    \label{eqn:TSS}
    \mathrm{TSS}=\frac{\mathrm{TP}}{\mathrm{TP}+\mathrm{FN}}-\frac{\mathrm{FP}}{\mathrm{FP}+\mathrm{TN}}\;,
\end{equation}
where $\mathrm{TP}$, $\mathrm{FP}$, $\mathrm{TN}$, and $\mathrm{FN}$ represent the number of true positives, false positives, true negatives, and false negatives, respectively. This metric is also used to guide parameter selection, ensuring that the score is maximized.

\subsection{Application to LASCO subset observations}
\label{subsec:expmultic}
We evaluated the classification performance on the dataset  previously described in Sect.~\ref{sec:exper}. According to Table~\ref{table:numberofevents}, the Super-wind $(\nearrow)$ class contains no events and is thus excluded from the following experiments. We applied the same data augmentation technique illustrated in Sect.~\ref{subsec:aug}, now extended to all non-empty speed-regime classes ($n=160\times 100$), while preserving the original class proportions.

\begin{figure*}
    \centering
    \includegraphics[width=\linewidth]{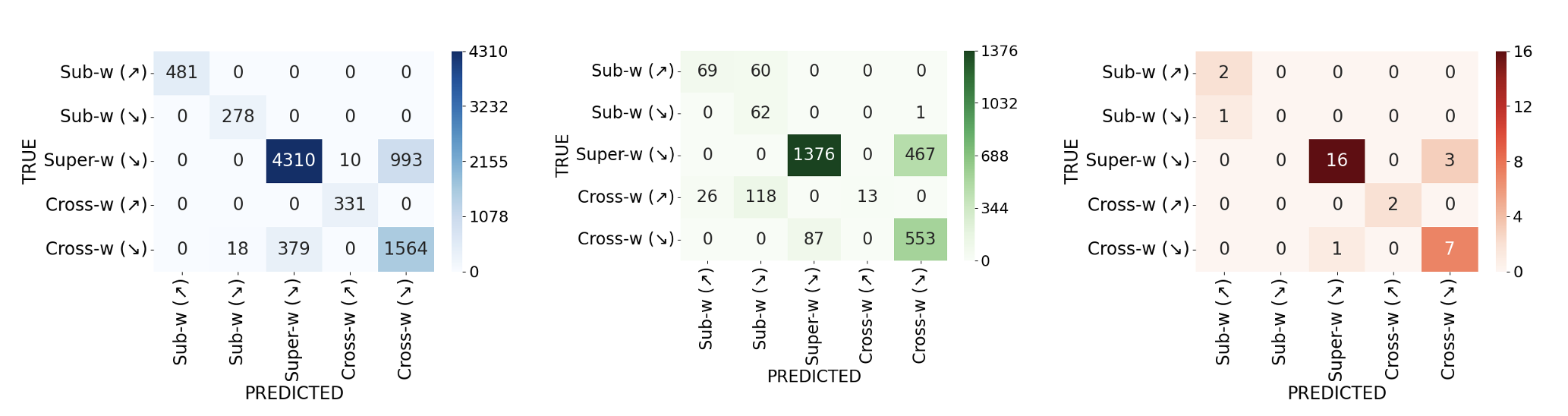}
    \caption{Confusion matrices for the multiclass classification results for training (left), validation (middle) and test (right) set of one of the 15 splits considered.}
    \label{fig:total}
\end{figure*}

Table~\ref{tab:splittotal} summarizes the average accuracy and TSS across 15 random splits for training, validation, and test sets, along with their standard deviations. Table~\ref{tab:testtotal} details the test performance (averaged over all splits) per class in a one-vs-all evaluation: for computing metrics, we treated the selected class as the positive class and all other classes combined as the negative class. This shows how well the model performs on each class separately, where each class was tested against all the others. In Fig.~\ref{fig:total}, the number of true positives and misclassifications are displayed in the form of confusion matrices for training, validation and test set of one representative split.
\begin{table}
\caption{Average metrics for the multiclass classification across all 15 splits.}
\centering
\begin{tabular}{lll}
\toprule\toprule
\textbf{Split} & \textbf{Accuracy} & \textbf{TSS} \\
\midrule
Train.  & 0.8391 $\pm$ 0.0185 & 0.8590 $\pm$ 0.0198 \\
Val.    & 0.7783 $\pm$ 0.0573 & 0.6328 $\pm$ 0.1179 \\
Test    & 0.7917 $\pm$ 0.0881 & 0.6287 $\pm$ 0.2101 \\
\bottomrule
\end{tabular}
\label{tab:splittotal}
\end{table}

\begin{table}
\caption{Average metrics for non-empty EDBM classes in the multiclass classification on the test set across all 15 splits.}
\centering
\begin{tabular}{lll}
\toprule\toprule
\textbf{Case} & \textbf{Accuracy} & \textbf{TSS} \\
\midrule
Sub-wind ($\nearrow$) & 0.9604 $\pm$ 0.0267 & 0.7089 $\pm$ 0.3102 \\
Sub-wind ($\searrow$) & 0.9625 $\pm$ 0.0306 & 0.4452 $\pm$ 0.5024 \\
Super-wind ($\searrow$) & 0.8458 $\pm$ 0.0640 & 0.6950 $\pm$ 0.1413 \\
Cross-wind ($\nearrow$)  & 0.9688 $\pm$ 0.0255 & 0.6556 $\pm$ 0.2976 \\
Cross-wind ($\searrow$) & 0.8458 $\pm$ 0.0564 & 0.6389 $\pm$ 0.1982 \\
\bottomrule
\end{tabular}
\tablefoot{The metrics were computed following a one-vs-all strategy (see text).}
\label{tab:testtotal}
\end{table}

Overall, the model achieves good generalization to unseen CME events (test-set accuracy around 79\% from Table~\ref{tab:splittotal}). The class imbalance impacts performance more noticeably in TSS than in accuracy, as reflected by the larger relative drop from the training set ($\mathrm{TSS}\approx 0.86$) to the validation and test sets ($\mathrm{TSS}\approx 0.63$). Class-wise, the model performs reliably, with a success rate above 84\% (Table~\ref{tab:testtotal}) and stable TSS values for the two most populated classes of the dataset (Super-wind $(\searrow)$ and Cross-wind $(\searrow)$). Less populated classes exhibit higher accuracy but larger variability in the TSS, a consequence of their limited sample sizes. This is particularly evident in the Sub-wind $(\searrow)$ class, the least populated: within the test set, this class, represented by only a single sample, was misclassified in 7 out of the 15 splits, resulting in an extremely large standard deviation. Fig.~\ref{fig:total} confirms that the classifier achieves robust separation among most regimes, although some confusion persists, especially for the Super-wind ($\searrow$) and Cross-wind ($\searrow$) classes. These misclassifications are more likely driven by intrinsic similarities in the CME kinematics of the two propagation regimes, both characterized by $v_0>v$, than by class imbalance.

\section{Conclusions and perspectives}
\label{sec:concl}
This work extends the applicability of AI frameworks based on kinematic models for travel-time prediction of CMEs traveling toward Earth by incorporating more physically informative CME propagation models. Integrating the closed-form solution of the EDBM \citep{rossi2025extended} directly into the neural-network loss function \citep{guastavino2023physics}, the approach achieves competitive predictions, with an MAE within 13 h. These results are comparable to, or improve upon, most existing CME forecasting methods (see Sect.~\ref{sec:intro}) and, importantly, are obtained for cataloged CME events that were previously excluded from standard DBM-based methods.

Introducing an extra, although necessary, unknown parameter in the physical model compared to the DBM, that is, the additional acceleration $a$, increases the model’s degrees of freedom. However, the ML workflow we designed overcomes this issue by enabling the AI model to extrapolate travel-time predictions for unseen CME events without explicitly computing $a$ during inference. The cost of this approach is transferred to the calibration of the drag coefficient $C$. A detailed sensitivity analysis revealed that the AI model is moderately sensitive to $C$ within a physically meaningful range, approximately around $C=100$. Values of 
$C$ outside this range, given the available dataset, led to underfitting or overfitting, and those far from this interval are associated with drag coefficients $\gamma$ that fall outside physically acceptable limits \citep{napoletano2022parameter,chierichini2024bayesian}.

In addition, we proposed an EDBM speed-regime multiclass classifier based on a multinomial logistic regression algorithm, capable of sorting CME events into their corresponding EDBM classes derived in Sect.~\ref{sec:edbm}, without requiring knowledge of the arrival CME speed $v$ during inference. Across multiple realizations, evaluation metrics indicate that the model generalizes well despite the limited number of events in certain speed classes, yielding consistent accuracy and stable error distributions. Nevertheless, the classification task becomes more challenging when the combined effects of class imbalance and CME dynamical similarities come into play.  

Extending and testing the present methodology for CME travel-time prediction to larger and (possibly) more balanced CME catalogs, encompassing all EDBM speed-regime classes, represents the next step toward developing an operationally reliable forecasting system. In this context, the multiclass classification module developed in this work serves as a key first stage in an operational pipeline, designed to automatically select the appropriate trained neural network, as discussed in Sect.~\ref{sec:multiclass}. Such a mechanism could form the backbone of future real-time CME forecasting systems, enabling both event classification and automatic model selection. 

\begin{figure*}
\tikzstyle{block} = [rectangle, rounded corners, text centered, draw=black, minimum width=1cm, minimum height=1cm]
\tikzstyle{decisionStyle} = [ellipse, draw, text centered, minimum width=1cm, minimum height=1cm]
\tikzstyle{arrow} = [thick, ->, >=stealth]
\centering
\begin{tikzpicture}[node distance=1.5cm, align=center]
\node (input) [block, fill=magenta!20] {\textbf{Input} \\ $\boldsymbol{x}=(v_0,m,A,\rho,w)$};
\node (class) [block, below=of input, fill=yellow!20] {\textbf{Speed-regime classification} \\ $\hat{c}\in\mathscr{C}=\{\text{Sub-wind ($\nearrow$)},\ldots,\text{Cross-wind ($\searrow$)}\}$};
\node (decision) [decisionStyle, below=of class, fill=orange!20] {$\hat{c} \in \{\text{Sub-wind ($\nearrow$)}, \text{Super-wind ($\searrow$)}\}$?};
\node (dbm) [block, below left=of decision, fill=cyan!20] {\textbf{Model selection} \\ DBM-trained neural network};
\node (edbm) [block, below right=of decision, fill=brown!20] {\textbf{Model selection} \\ EDBM-trained neural network};
\node (output) [block, below =of decision, yshift=-2cm, fill=green!20] {\textbf{Arrival time at 1 AU} \\ $\hat{t}$ [h]};
\draw [arrow] (input) -- (class) node[midway, right] {Classifier};
\draw [arrow] (class) -- (decision);
\draw [arrow] (decision) -| node[midway, above]{YES} (dbm);
\draw [arrow] (decision) -| node[midway, above]{NO} (edbm);
\draw [arrow] (dbm) |- node[midway, below]{$\mathcal{NN}_{\text{DBM}}$} (output);
\draw [arrow] (edbm) |- node[midway, below]{$\mathcal{NN}_{\text{EDBM},\hat{c}}$} (output);
\end{tikzpicture}
\caption{Schematic flow chart of the proposed pipeline for operational CME travel-time forecasting. $\mathcal{NN}_{\text{DBM}}$ and $\mathcal{NN}_{\text{EDBM},\hat{c}}$ denote the two model-specific neural networks used to produce the predicted time $\hat{t}$, with the EDBM network conditioned on the predicted speed-regime class $\hat{c}$.}
\label{fig:schem}
\end{figure*}
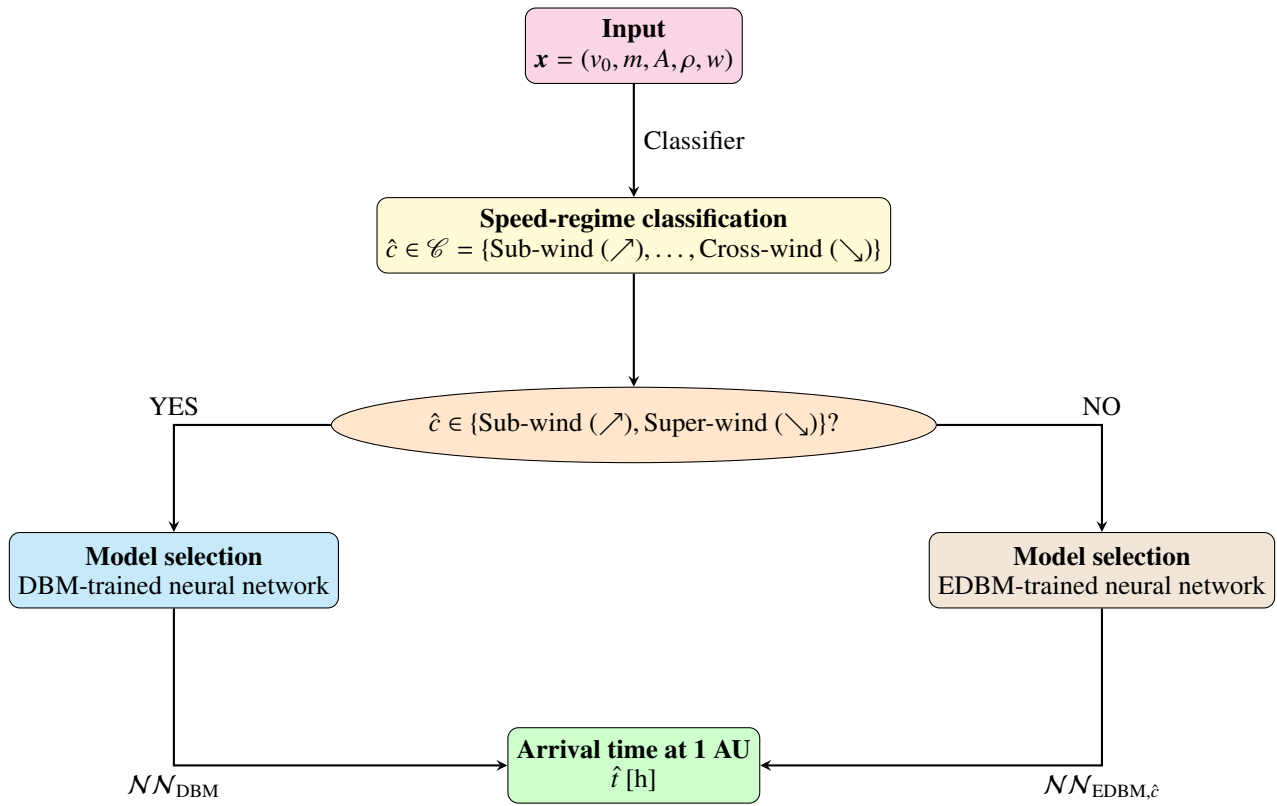

The envisioned pipeline would systematically address two core challenges from new observational data:
\begin{enumerate}
    \item automatic classification of CME events to identify the appropriate dynamical regime;
    \item travel-time prediction.
\end{enumerate}
For DBM-compatible propagation regimes, both the EDBM-based and DBM-based trained neural networks (the latter from \citealp{guastavino2023physics}) could be applied. When both models are applicable, a comparative performance analysis will be essential to identify the more accurate network for inference. Fig.~\ref{fig:schem} shows a schematic of the proposed operational workflow, in which, for simplicity, the DBM is assumed to be preferred whenever applicable. 
As more extensive and diverse CME datasets become available in the coming years, future efforts will focus on refining and validating this ambitious integrated framework toward real-time, physics-informed ML-based space weather forecasting.

\begin{acknowledgements}\\
All authors acknowledge the support of the Fondazione Compagnia di San Paolo within the framework of the Artificial Intelligence Call for Proposals, AIxtreme project (ID Rol: 71708). ML, SG, MP and AMM are also grateful to the Gruppo Nazionale per il Calcolo Scientifico - Istituto Nazionale di Alta Matematica (GNCS - INdAM). MR is also grateful to the Gruppo Nazionale per la Fisica Matematica -- Istituto Nazionale di Alta Matematica (GNFM -- INdAM).
\end{acknowledgements}

\bibliographystyle{aa} 
\bibliography{references}

\end{document}